\documentclass[conference]{IEEEtran}
\IEEEoverridecommandlockouts

\usepackage{amssymb}
\usepackage[cmex10]{amsmath}
\usepackage[bottom=1in,top=0.8in,left=0.7in,right=0.7in]{geometry}
\usepackage{stfloats}
\usepackage{subfigure}
\usepackage{epsfig,epsf,color,balance,cite}
\usepackage{bm}
\usepackage{tikz}

\usepackage{algorithm}
\usepackage{algorithmic}

\usepackage{color}
\definecolor{myc1}{rgb}{0,0,0}

\begin{document}

\title{{An Efficient Relay Selection Scheme for Relay-assisted HARQ
}  }

\author{
\IEEEauthorblockN{Weihang Ding\IEEEauthorrefmark{1} and Mohammad Shikh-Bahaei\IEEEauthorrefmark{1}
                  }
\IEEEauthorblockA{\IEEEauthorrefmark{1}Centre for Telecommunications Research, Department of Engineering, King's College London, London WC2R 2LS, UK.}

\vspace{-2em}
}

\maketitle

\begin{abstract}
In wireless communication networks, relays are required when the quality of the direct link between the source and the destination is not high enough to support reliable transmission because of long distances or obstacles. Selecting the proper relay node (RN) to support hybrid automatic repeat request (HARQ) is of great importance in such a relay-assisted network. Different from previous works, whether to participate in the transmission is determined by each RN itself in this work, thus reducing the overhead. As RNs do not need to obtain channel state information about the whole network, there is no significant overhead in the system. Using the numbers of transmission attempts required by both channels calculated from the obtained channel state information, each RN sets a timer and forwards the packet when time-out occurs.
Simulation results show that our proposed method significantly improves the performance of the system. When the channels are of relatively high quality, the performance of our method is close to the optimal relay selection which requires full information about the network.

\end{abstract}

\IEEEpeerreviewmaketitle

\section{Introduction}

Relay-assisted communication is a promising technique used in wireless networks. If the direct link from the source to the destination is not good enough to complete the transmission, the assistance of relay nodes (RNs) can significantly improve the efficiency and robustness of the network. 
RNs not only improve coverage and connectivity but also provide spatial diversity.
In \cite{Laneman}, two relaying protocols namely amplify-and-forward (AF), and decode-and-forward (DF) are proposed to improve the overall performance. Compared with AF protocol, which only amplifies the received signal\cite{Yuksel}, DF protocol is more efficient as noise is not accumulated at the RNs \cite{Bhatnagar, Bhatnagar2}. 

Hybrid automatic repeat request (HARQ) is a feasible solution to improve the reliability of relay networks\cite{J.Kim}.
Combining automatic repeat request (ARQ) and error-correcting codes, HARQ is a well-established technique handling retransmissions at the physical and MAC layer for reliable transmissions over fading channels\cite{Wicker}.
In terms of the soft-combining method, HARQ protocols can be categorized into chase-combining HARQ (CC-HARQ) and incremental redundancy HARQ (IR-HARQ)\cite{Mandelbaum}. In CC-HARQ, the same codewords are transmitted in different transmission attempts while in IR-HARQ, different codewords are transmitted in the retransmissions to lower the code rate. These two protocols are investigated in \cite{Caire} and it is shown that IR-HARQ can approach the ergodic channel capacity over block fading channels. The performance can be further enhanced by making the system adaptive \cite{2,3,4,6,10,12}.

Plenty of works have been performed on optimizing the performance of HARQ in relay networks \cite{12,13,14,15}. Energy efficiency optimization is performed in \cite{Shuguang, Maaz, J.Choi, S.Lee}. In \cite{J.Choi, Chelli,5}, the average delay of such networks is studied from an information theoretic perspective. The throughput or spectral efficiency is optimized in \cite{Chelli, Han2017SpectralEO}. Network coding \cite{Tutgun} and MIMO \cite{Bhatnagar2013} are applied for improving the reliability of the system.
In reality, there are usually multiple potential RNs in the network. In \cite{Hu2016}, an RN participates in the transmission as long as it recovers the message earlier than the destination. However, due to the limits of the resources \cite{1,11}, it is shown in \cite{Beres} that finding one best RN is more beneficial than deploying several RNs simultaneously.
In \cite{J.Kim}, the author proposes a method called transmission number relaying (TNR) for RN selection given the assumption that each RN in the network can obtain full information about the network from the negative acknowledgment (i.e., NACK), which results in extra overhead.
In \cite{Q.Vien}, two efficiency matrices are formulated for relay selection to minimize the transmission delay and energy consumption where network coding is applied.
In \cite{X.Du}, deep reinforcement learning is used for RN selection in Long Term Evolution vehicle-to-everything communication (LTE-V) without global information.

In this work, we focus on the IR-HARQ protocol in a wireless network containing multiple RNs.
We propose a RN selection scheme based on the qualities of the source-RN and RN-destination channels. Upon receiving the feedback from the destination which indicates whether the assistance of RNs is required, each RN sets up a timer and starts forwarding the packet as soon as the timer expires. The timer is set separately and independently by each RN itself according to an algorithm, hence, no extra overhead is introduced.

\section{System model and the relay selection scheme}

\begin{figure}[t!]
    \centering
    \includegraphics[width=0.7\columnwidth]{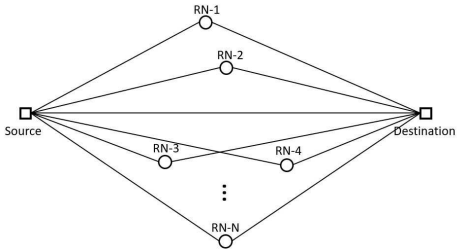}
    \caption{The relay network with $N$ RNs.}
    \label{system model}
\end{figure}

We consider a network with one source, one destination, and $N$ RNs, all operating in half-duplex mode (See Fig. \ref{system model}). The direct link between the source and the destination does exist but is not always good enough to ensure successful transmission.

If the source-destination link is not good enough, RNs are required to assist the transmission by forwarding the received packets to the destination.
We apply IR-HARQ in this network to improve the reliability of the system.
To limit the average delay of a HARQ process, the maximum number of transmission attempts allowed in a single HARQ round $M$ is always set as a relatively small value. After each transmission attempt, the destination generates feedback to inform the sender of the decodability of the codeword. If the transmitter observes an ACK, the current HARQ round will be terminated. 
However, if a NACK is received, the source will continue transmitting redundancy bits in the subsequent transmission until it finally receives an ACK.
The feedback channel is assumed to be perfect, so an outage will only occur when:
\begin{itemize}
    \item The source-destination channel is not good enough to transmit the packet successfully within $M$ transmissions.
    \item There are no such RNs, that the numbers of transmissions required by both the source-RN and the RN-destination links do not exceed $M$.
\end{itemize}

The channels between any two nodes are assumed to be Rayleigh fading channel which varies relatively slowly. Therefore, each channel remains constant during the whole HARQ round.
We denote the channel gains between the source and the $k$-th RN, and between the $k$-th RN and the destination as $h_{s,k}$ and $h_{k,d}$ respectively. 
For simplicity, we assume that the average channel gains of different channels are equal, hence $\mathbb{E}[|h_{s,k}|^2]=\mathbb{E}[|h_{k,d}|^2]=\bar{\gamma}, k=1,\dots,N$.

Both the source and the RNs can only transmit with fixed power $P$. The noise of each channel is Gaussian distributed with zero mean and variance $\sigma^2$. So, the signal-to-noise ratio of all the channels can be written as SNR$=P/\sigma^2$. In this system, we also assume that every two nodes are placed within the single-hop transmission range, and only the receiver can obtain the channel state information from the transmitted signal.
The received signal at the $k$-th RN from the source, and at the destination from the $k$-th RN can be respectively modeled as:

\begin{equation}
    \bm{y_{s,k}}=\sqrt{P}h_{s,k}\bm{x}+\bm{n_{s,k}},
\end{equation}
\begin{equation}
    \bm{y_{k,d}}=\sqrt{P}h_{k,d}\bm{x}+\bm{n_{k,d}},
\end{equation}
where $\bm{x}$ is the input signal, and $\bm{n_{s,k}}$ $\bm{n_{k,d}}$ are the noise of all three channels respectively.

\begin{figure}[t!]
    \centering
    \includegraphics[width=0.85\columnwidth]{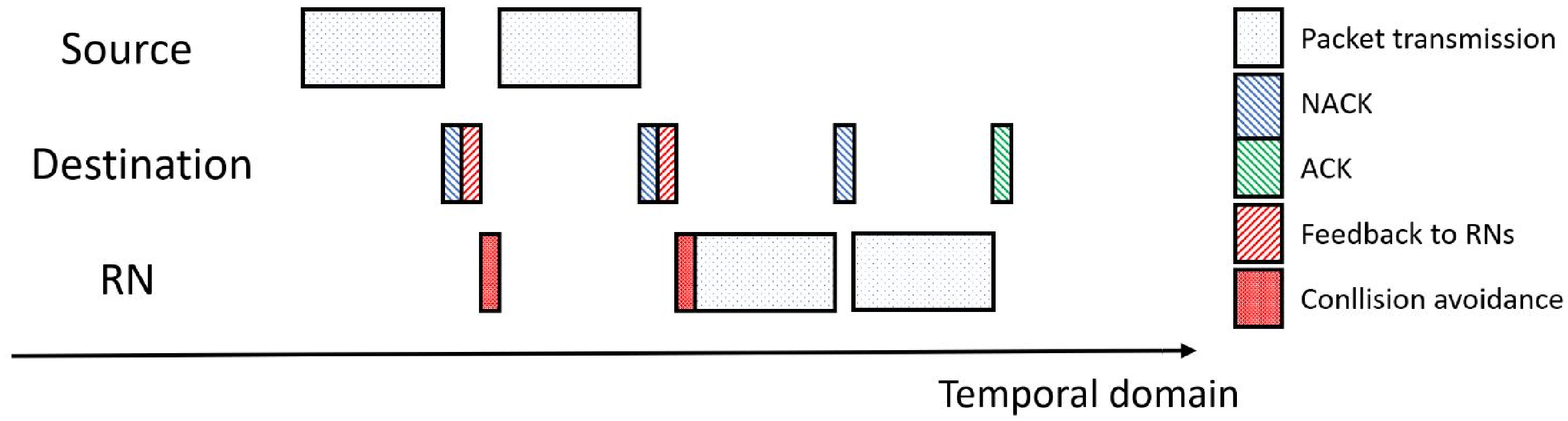}
    \caption{An example of a HARQ round in temporal domain: The source-destination channel is not good enough to support the transmission. This message is transmitted to all RNs by extending the NACK by 1 bit. After the second transmission attempt, the timer at a RN expires and it takes over the transmission. After four transmissions in total, the destination eventually decodes the packet and notifies the RNs and the source with an ACK.}
    \label{time domain}
\end{figure}

In the initial transmission of each HARQ round, the destination and all the RNs receive the packet from the source and save it in the buffer. 
Using the channel state information obtained from the received packet, all the RNs predict how many transmission attempts are required. If the required number of transmission attempts exceeds $M$, this relay will clear the buffer and stops listening to save energy.
In the meantime, the destination attempts to decode the packet. If the message is successfully recovered, the destination will broadcast an ACK, otherwise, it will predict whether the assistance of RNs is required and append this message to the NACK. If the feedback indicates that RN assistance is unnecessary, the RNs will clear their buffers as well.
If RN assistance is required, the RNs will predict the number of transmission attempts required by the relay-destination channel, set a timer, and keep listening to the other RNs. All the RNs transmit the packet using the same resource block.
If the timer at a RN expires, and the resource block is still idle, this RN takes over the transmission and starts forwarding the packet to the destination.
If the RN detects that the resource block is occupied, it prohibits the transmission, clears the buffer, and waits for subsequent data packets.
The procedures are shown in Fig. \ref{time domain} in temporal domain when relay assistance is necessary. A short period of time is reserved for collision avoidance, which will be discussed in the next section.

The transmission will succeed as long as at least one RN can successfully relay the packet before outage occurs.
Choosing different RNs does not lead to any changes in outage probability, so we only consider the HARQ performance without outages in this work.

\section{Problem formulation}

To simplify the system, we assume that the signal-to-noise ratios of all channels are equal, but the channel gains are still variable due to fading.
The decodability of a packet is reflected by the accumulated mutual information (MI). The received MI of each resource symbol at the $k$-th relay is given by:
\begin{equation}
    I_{s,k}=\log_2(1+\text{SNR}|h_{s,k}|^2).
\end{equation}

Similarly, the received MI of each resource symbol at the destination from the $k$-th relay is given by:
\begin{equation}
    I_{k,d}=\log_2(1+\text{SNR}|h_{k,d}|^2).
\end{equation}

If the initial transmission cannot be recovered, the received packet is not discarded, but saved in a buffer and soft-combined with the following transmissions.
The accumulated MI of IR-HARQ is the summation of MI in each transmission \cite{Jabi}. 
The transmission will not terminate until the transmitter receives a message indicating successful decoding at the receiver or detects that the bandwidth reserved for the RNs is occupied.
Since the packet length and transmission power of each transmission attempt remain constant in a HARQ round, the minimum number of transmissions required to ensure a successful packet delivery via the channel between the source and the $k$-th RN $N_{s,k}$ can be predicted by the RN itself using the channel gain.

The decodability of a packet can be reflected by whether the accumulated MI is greater than the spectral efficiency of the original message. Therefore, $N_{s,k}$ transmissions are required for a successful transmission from the source to the $k$-th RN if the accumulated MI after the $(N_{s,k}-1)$-th transmission is lower than the spectral efficiency $R$ and the accumulated MI after the $N_{s,k}$-th transmission is greater than $R$.
Similarly, $N_{k,d}$ is the number of transmissions required by the channel between the $k$-th RN and the destination. 
$N_{s,k}$ and $N_{k,d}$ can be calculated as follows:

\begin{equation}
    N_{s,k} = \lceil\frac{R}{\log_2(1+\text{SNR}|h_{s,k}|^2)} \rceil,
\end{equation}

\begin{equation}
    N_{k,d} = \lceil\frac{R}{\log_2(1+\text{SNR}|h_{k,d}|^2)} \rceil.
\end{equation}

The probability that $N_{s,k}=i$ is denoted as $P_{sk,i}$, which can be written as:
\begin{equation}
    \begin{aligned}
    P_{sk,i}&=\mathbb{P}\{(i-1)\log_2(1+\text{SNR}|h_{s,k}|^2)<R\\
    &\:\:\:\:\:<i\log_2(1+\text{SNR}|h_{s,k}|^2) \}\\
    &=\mathbb{P}\left\{\frac{2^{R/(i-1)}-1}{\text{SNR}}<|h_{s,k}|^2<\frac{2^{R/i}-1}{\text{SNR}} \right\}\\
    &=\exp{\left(-\frac{2^{R/i}-1}{\bar{\gamma}\text{SNR}}\right)}-\exp{\left(-\frac{2^{R/(i-1)}-1}{\bar{\gamma}\text{SNR}}\right)}.\\
    \end{aligned}
\end{equation}

As the channel gains of different channels are independently distributed, the joint probability that $N_{s,k}=i$, and $N_{k,d}=j$ can be calculated as:
\begin{equation}
\begin{aligned}
    P_{i,j}=&P_{sk,i}\times P_{kd,j}\\
    =&\left(\exp\left(-\frac{2^{R/i}-1}{\bar{\gamma}\text{SNR}}\right)-\exp\left(-\frac{2^{R/(i-1)}-1}{\bar{\gamma}\text{SNR}}\right)\right)\\
    &\times\left(\exp\left(-\frac{2^{R/j}-1}{\bar{\gamma}\text{SNR}}\right)-\exp\left(-\frac{2^{R/(j-1)}-1}{\bar{\gamma}\text{SNR}}\right)\right).
\end{aligned}
\end{equation}
where $P_{kd,j}$ is the probability that $N_{k,d}=j$.

Based on $i$ and $j$, each RN sets the timer to a value $T_{i,j}$, which is calculated as:
\begin{equation}
    T_{i,j}=\alpha_{i,j} T_{tr}+\beta,
\label{timer}
\end{equation}
where $T_{tr}$ is the time interval between each transmission and the previous transmission, which remains constant and is known prior to the transmission. $\alpha_{i,j}\in [1,M]$ is an integer determined by $N_{s,k}$ and $N_{k,d}$.
$\beta$ is a randomly chosen value irrelevant of $N_{s,k}$ and $N_{k,d}$, which is much smaller than $T_{tr}$. Different $\beta$ values are distributed to different RNs for collision avoidance because all the RNs will forward the packet using the same resource block.

The performance of the HARQ process can be reflected by the average delay and the throughput. Here, the throughput is defined as the number of information bits that each transmitted symbol can successfully deliver in a whole HARQ round.
We aim to minimize the average delay and maximize the overall throughput by selecting proper $\bm{\alpha}$, which is a matrix shown as follows:
\begin{equation}
\bm{\alpha}=
\begin{bmatrix}
\alpha_{1,1} & \dots & \alpha_{1,M}\\
\vdots & \ddots & \vdots\\
\alpha_{M,1} & \dots &\alpha_{M,M}
\end{bmatrix}.
\end{equation}

All the possible $(i,j)$ pairs are divided into $M$ sets $\mathbb{S}_1,\dots,\mathbb{S}_M$ based on the values of $\alpha_{i,j}$. $\mathbb{S}_n$ contains all the $(i,j)$ pairs satisfying $\alpha_{i,j}=n$.
We define $S_1, \dots, S_M$ as the events that there is at least one relay nodes whose $(i,j)$ is within $\mathbb{S}_1, \dots, \mathbb{S}_M$, respectively.
The average delay can be calculated as:
\begin{equation}
\begin{aligned}
\mathbb{E}[D]=&\left(\sum_{n=1}^{M} \frac{\sum_{(i,j)\in\mathbb{S}_n}(n+j)P_{i,j}}{\sum_{(i,j)\in\mathbb{S}_n}P_{i,j}}\times \mathbb{P}\{S_n\cap \prod_{k=0}^{n-1}\bar{S_k}\}\right)\\
&/\mathbb{P}\{S_1\cup \dots \cup S_M\},
\end{aligned}
\label{op_delay}
\end{equation}
where $S_0$ contains all $(i,j)$ pairs. The throughput of the same system $\eta$ depends on the total number of transmission attempts given that the length of each transmission is the same, hence can be written as:
\begin{equation}
    \eta=\frac{\mathbb{P}\{S_1\cup \dots \cup S_M\}}{\left(\sum_{n=1}^{M} \frac{\sum_{(i,j)\in\mathbb{S}_n}(i+j)P_{i,j}}{\sum_{(i,j)\in\mathbb{S}_n}P_{i,j}}\times \mathbb{P}\{S_n\cap \prod_{k=0}^{n-1}\bar{S_k}\}\right)\times R}.
    \label{op_th}
\end{equation}

Based on these two formulas above, we can simply formulate two optimization problems optimizing the average delay and throughput separately:
\begin{equation}
    \begin{aligned}
        \min_{\bm{\alpha}}\:&\mathbb{E}[D]\\
        \rm{subject\:to:}\:\:& \alpha_{i,j}\in \mathbb{R}\\
        &\alpha_{i,j}\leq M
    \end{aligned}
\end{equation}

\begin{equation}
    \begin{aligned}
        \max_{\bm{\alpha}}\:&\eta\\
        \rm{subject\:to:}\:\:& \alpha_{i,j}\in \mathbb{R}\\
        &\alpha_{i,j}\leq M.
    \end{aligned}
    \label{op2}
\end{equation}

These two optimization problems are actually hard to solve. However, $M$ is usually a small number, hence we can solve them using an iterative algorithm. The algorithm that is used for delay minimization is shown as Alg.\ref{alg}. $\bm{\alpha}$ is initialized such that $\alpha_{i,j}=i$. In each iteration, one value in the matrix is optimized and updated while the others are fixed.
As a better $\bm{\alpha}$ is obtained after each iteration, we will eventually find out the global optimization.
This algorithm terminates when $\bm{\alpha}$ remains unchanged within two consecutive iterations.
Using this algorithm, we can obtain the optimal policy for each $(i,j)$ pair, and it is not very time-costly. Similarly, (\ref{op2}) can be solved by replacing (\ref{op_delay}) with (\ref{op_th}) in step 12 and selecting $k$ corresponding to the great value in the vector in step 15 of Alg.\ref{alg}.

\begin{algorithm}[t!]
\caption{$\bm{\alpha}$ determination}

\begin{algorithmic}[1]

\STATE Set the initial $\bm{\alpha}$: $\alpha_{i,j}=i, \forall i,j \in [1,M]$.
\REPEAT
\STATE Set $i=1, j=1$.
\FOR{$i\leq M$}
    \FOR{$j\leq M$}
    \STATE Set $k=1$.
        \FOR{$k\leq M$}
        \IF{$k<i$}
        \STATE Set $E_k=\infty$
        \ELSE
        \STATE Replace $\alpha_{i,j}$ with $k$, and calculate $\mathbb{E}[D]$ using (\ref{op_delay}).
        \STATE Set $E_k=\mathbb{E}[D]$.
        \ENDIF
        \ENDFOR
        \STATE Compare $E_1,\dots,E_M$, find the $k$ that minimizes $E_k$, update $\alpha_{i,j}=k$.
    \ENDFOR
\ENDFOR
\UNTIL{$\bm{\alpha}$ remains unchanged after a loop.}
\STATE Output $\alpha$.

\end{algorithmic}
\label{alg}

\end{algorithm}

\section{Numerical results}

In this section, the performance of our proposed method is compared with other existing RN selection schemes. We assume that the maximum number of transmission attempts of each link $M=4$, and the spectral efficiency of the initial transmission $R=1$ bit/symbol. First decode, first relay (FDFR) scheme refers to the scenario where each relay directly forwards the packet after successful decoding as long as the number of transmissions required by the channel between the destination and itself is less than $M$. Optimal relay selection is simply selecting the RN that minimizes the summation of transmissions required by both links. Optimal relay selection requires complete information on all channels in the relay network, which leads to more overhead. In the comparison, the effect of $\beta$ in (\ref{timer}) can be ignored because it is rather small compared with $T_{tr}$.

\begin{figure}[t!]
    \centering
    \includegraphics[width=0.95\columnwidth]{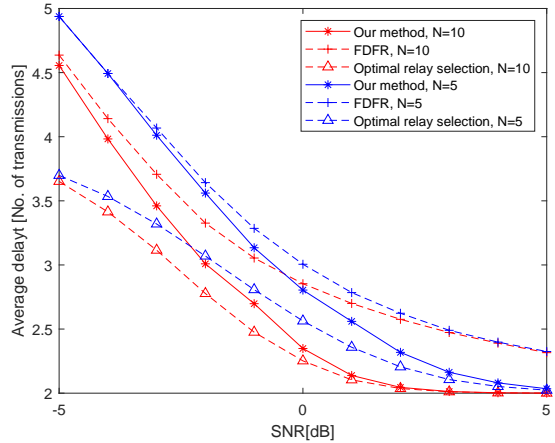}
    \caption{The average delay of this relay-assisted HARQ process at different SNR values for $N=5$, and $N=10$.}
    \label{delay}
\end{figure}

\begin{figure}[t!]
    \centering
    \includegraphics[width=0.95\columnwidth]{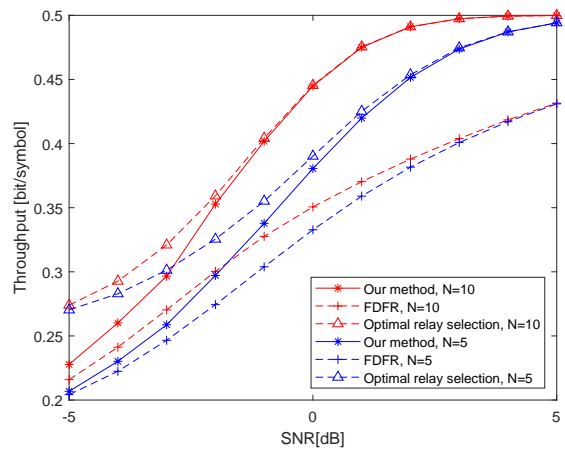}
    \caption{The throughput of this system when the number of users in the network $N=5$, and $N=10$ respectively.}
    \label{throughput}
\end{figure}

From Fig. \ref{delay}, we can see that the average delay (in terms of the number of transmission attempts) of optimal relay selection is far lower than FDFR, which matches our expectations.
The delay of our proposed scheme is slightly lower than the FDFR scheme in low SNR regimes and approaches optimal relay selection when the qualities of the channels are high. We can also see that when there are more available RNs in the network, the average delay is lower in low SNR regimes, and our method converges to optimal relay selection at lower SNR values. This is because more RNs and better channel qualities can both reduce the uncertainty of the network.
The throughputs of the same network are compared in Fig. \ref{throughput} for $N=5$ and $N=10$ respectively. Similar to Fig. \ref{delay}, the throughput of optimal relay selection is much higher than FDFR, and our method lie between them. 
Due to inappropriate RN selection, the throughput of FDFR suffers significantly compared to optimal relay selection, and our proposed scheme can mitigate this impact.

\section{conclusion}
In this paper, we propose a novel RN selection scheme to select the sub-optimal RN that is used for supporting IR-HARQ in the wireless network. The selections are sub-optimal because all the RNs make decisions without full information about all the channels in the network, which significantly reduces the overhead. The RN selection is carried out by the RN itself and the algorithm is not complicated. Numerical results show that the performance of our proposed method in terms of the average relay and overall throughput is close to the optimal RN selection at higher SNR regimes.


\vspace{-0.5em}
\bibliographystyle{IEEEtran}
\bibliography{IEEEabrv,MMM}

\begin{thebibliography}{10}
\providecommand{\url}[1]{#1}
\csname url@samestyle\endcsname
\providecommand{\newblock}{\relax}
\providecommand{\bibinfo}[2]{#2}
\providecommand{\BIBentrySTDinterwordspacing}{\spaceskip=0pt\relax}
\providecommand{\BIBentryALTinterwordstretchfactor}{4}
\providecommand{\BIBentryALTinterwordspacing}{\spaceskip=\fontdimen2\font plus
\BIBentryALTinterwordstretchfactor\fontdimen3\font minus
  \fontdimen4\font\relax}
\providecommand{\BIBforeignlanguage}[2]{{%
\expandafter\ifx\csname l@#1\endcsname\relax
\typeout{** WARNING: IEEEtran.bst: No hyphenation pattern has been}%
\typeout{** loaded for the language `#1'. Using the pattern for}%
\typeout{** the default language instead.}%
\else
\language=\csname l@#1\endcsname
\fi
#2}}
\providecommand{\BIBdecl}{\relax}
\BIBdecl

\bibitem{Laneman}
J.~N. {Laneman}, D.~N.~C. {Tse}, and G.~W. {Wornell}, ``Cooperative diversity
  in wireless networks: Efficient protocols and outage behavior,'' \emph{IEEE
  Transactions on Information Theory}, vol.~50, no.~12, pp. 3062--3080, 2004.

\bibitem{Yuksel}
M.~{Yuksel} and E.~{Erkip}, ``Diversity in relaying protocols with amplify and
  forward,'' in \emph{GLOBECOM '03. IEEE Global Telecommunications Conference
  (IEEE Cat. No.03CH37489)}, vol.~4, 2003, pp. 2025--2029 vol.4.

\bibitem{Bhatnagar}
M.~R. {Bhatnagar} and A.~{Hjorungnes}, ``Ml decoder for decode-and-forward
  based cooperative communication system,'' \emph{IEEE Transactions on Wireless
  Communications}, vol.~10, no.~12, pp. 4080--4090, 2011.

\bibitem{Bhatnagar2}
M.~R. {Bhatnagar}, ``On the capacity of decode-and-forward relaying over rician
  fading channels,'' \emph{IEEE Communications Letters}, vol.~17, no.~6, pp.
  1100--1103, 2013.

\bibitem{J.Kim}
J.~{Kim}, K.~{Kim}, and J.~{Lee}, ``Energy-efficient relay selection of
  cooperative harq based on the number of transmissions over rayleigh fading
  channels,'' \emph{IEEE Transactions on Vehicular Technology}, vol.~66, no.~1,
  pp. 610--621, 2017.

\bibitem{Wicker}
S.~B. Wicker, \emph{Error Control Systems for Digital Communication and
  Storage}.\hskip 1em plus 0.5em minus 0.4em\relax USA: Prentice-Hall, Inc.,
  1994.

\bibitem{Mandelbaum}
D.~{Mandelbaum}, ``An adaptive-feedback coding scheme using incremental
  redundancy (corresp.),'' \emph{IEEE Transactions on Information Theory},
  vol.~20, no.~3, pp. 388--389, 1974.

\bibitem{Caire}
G.~{Caire} and D.~{Tuninetti}, ``The throughput of hybrid-arq protocols for the
  gaussian collision channel,'' \emph{IEEE Transactions on Information Theory},
  vol.~47, no.~5, pp. 1971--1988, 2001.

\bibitem{2}
H.~Bobarshad, M.~van~der Schaar, and M.~R. Shikh-Bahaei, ``A low-complexity
  analytical modeling for cross-layer adaptive error protection in video over
  wlan,'' \emph{IEEE Transactions on Multimedia}, vol.~12, no.~5, pp. 427--438,
  2010.

\bibitem{3}
V.~Towhidlou and M.~Shikh-Bahaei, ``Adaptive full-duplex communications in
  cognitive radio networks,'' \emph{IEEE Transactions on Vehicular Technology},
  vol.~67, no.~9, pp. 8386--8395, 2018.

\bibitem{4}
M.~Shikh-Bahaei, ``Joint optimization of “transmission rate” and
  “outer-loop snr target” adaptation over fading channels,'' \emph{IEEE
  Transactions on Communications}, vol.~55, no.~3, pp. 398--403, 2007.

\bibitem{6}
K.~Nehra and M.~Shikh-Bahaei, ``Spectral efficiency of adaptive mqam/ofdm
  systems with cfo over fading channels,'' \emph{IEEE Transactions on Vehicular
  Technology}, vol.~60, no.~3, pp. 1240--1247, 2011.

\bibitem{10}
Y.~Zhang, J.~Hou, V.~Towhidlou, and M.~R. Shikh-Bahaei, ``A neural network
  prediction-based adaptive mode selection scheme in full-duplex cognitive
  networks,'' \emph{IEEE Transactions on Cognitive Communications and
  Networking}, vol.~5, no.~3, pp. 540--553, 2019.

\bibitem{12}
Y.~Zhang, Q.~Wu, and M.~R. Shikh-Bahaei, ``On ensemble learning-based secure
  fusion strategy for robust cooperative sensing in full-duplex cognitive radio
  networks,'' \emph{IEEE Transactions on Communications}, vol.~68, no.~10, pp.
  6086--6100, 2020.

\bibitem{13}
V.~Towhidlou and M.~Shikh-Bahaei, ``Improved cognitive networking through full
  duplex cooperative arq and harq,'' \emph{IEEE Wireless Communications
  Letters}, vol.~7, no.~2, pp. 218--221, 2018.

\bibitem{14}
------, ``Cooperative arq in full duplex cognitive radio networks,'' in
  \emph{2016 IEEE 27th Annual International Symposium on Personal, Indoor, and
  Mobile Radio Communications (PIMRC)}, 2016, pp. 1--5.

\bibitem{15}
A.~Kobravi and M.~Shikh-Bahaei, ``Cross-layer adaptive arq and modulation
  tradeoffs,'' in \emph{2007 IEEE 18th International Symposium on Personal,
  Indoor and Mobile Radio Communications}, 2007, pp. 1--5.

\bibitem{Shuguang}
{Shuguang Cui}, A.~J. {Goldsmith}, and A.~{Bahai}, ``Energy-constrained
  modulation optimization,'' \emph{IEEE Transactions on Wireless
  Communications}, vol.~4, no.~5, pp. 2349--2360, 2005.

\bibitem{Maaz}
M.~{Maaz}, P.~{Mary}, and M.~{Hélard}, ``Energy minimization in harq-i
  relay-assisted networks with delay-limited users,'' \emph{IEEE Transactions
  on Vehicular Technology}, vol.~66, no.~8, pp. 6887--6898, 2017.

\bibitem{J.Choi}
J.~{Choi}, W.~{Xing}, D.~{To}, Y.~{Wu}, and S.~{Xu}, ``On the energy efficiency
  of a relaying protocol with harq-ir and distributed cooperative
  beamforming,'' \emph{IEEE Transactions on Wireless Communications}, vol.~12,
  no.~2, pp. 769--781, 2013.

\bibitem{S.Lee}
S.~{Lee}, W.~{Su}, S.~{Batalama}, and J.~D. {Matyjas}, ``Cooperative
  decode-and-forward arq relaying: Performance analysis and power
  optimization,'' \emph{IEEE Transactions on Wireless Communications}, vol.~9,
  no.~8, pp. 2632--2642, 2010.

\bibitem{Chelli}
A.~{Chelli}, E.~{Zedini}, M.~{Alouini}, M.~{Pätzold}, and I.~{Balasingham},
  ``Throughput and delay analysis of harq with code combining over double
  rayleigh fading channels,'' \emph{IEEE Transactions on Vehicular Technology},
  vol.~67, no.~5, pp. 4233--4247, 2018.

\bibitem{5}
H.~Bobarshad and M.~Shikh-Bahaei, ``M/m/1 queuing model for adaptive
  cross-layer error protection in wlans,'' in \emph{2009 IEEE Wireless
  Communications and Networking Conference}, 2009, pp. 1--6.

\bibitem{Han2017SpectralEO}
J.~Han, Y.~Xi, and J.~Wei, ``Spectral efficient optimisation for multihop relay
  networks adopting harq schemes,'' \emph{Electronics Letters}, vol.~53, pp.
  358--360, 2017.

\bibitem{Tutgun}
R.~{Tutgun} and E.~{Aktas}, ``Cooperative network coded arq strategies for
  two-way relay channels,'' \emph{IEEE Transactions on Vehicular Technology},
  vol.~64, no.~7, pp. 3205--3217, 2015.

\bibitem{Bhatnagar2013}
M.~R. {Bhatnagar} and M.~K. {Arti}, ``Selection beamforming and combining in
  decode-and-forward mimo relay networks,'' \emph{IEEE Communications Letters},
  vol.~17, no.~8, pp. 1556--1559, 2013.

\bibitem{Hu2016}
Y.~{Hu}, J.~{Gross}, and A.~{Schmeink}, ``Qos-constrained energy efficiency of
  cooperative arq in multiple df relay systems,'' \emph{IEEE Transactions on
  Vehicular Technology}, vol.~65, no.~2, pp. 848--859, 2016.

\bibitem{1}
Z.~Yang, J.~Hou, and M.~Shikh-Bahaei, ``Energy efficient resource allocation
  for mobile-edge computation networks with noma,'' in \emph{2018 IEEE Globecom
  Workshops (GC Wkshps)}, 2018, pp. 1--7.

\bibitem{11}
A.~Shadmand and M.~Shikh-Bahaei, ``Multi-user time-frequency downlink
  scheduling and resource allocation for lte cellular systems,'' in \emph{2010
  IEEE Wireless Communication and Networking Conference}, 2010, pp. 1--6.

\bibitem{Beres}
E.~{Beres} and R.~{Adve}, ``Selection cooperation in multi-source cooperative
  networks,'' \emph{IEEE Transactions on Wireless Communications}, vol.~7,
  no.~1, pp. 118--127, 2008.

\bibitem{Q.Vien}
Q.~{Vien}, H.~X. {Nguyen}, P.~{Shah}, E.~{Ever}, and D.~{To}, ``Relay selection
  for efficient harq-ir protocols in relay-assisted multisource multicast
  networks,'' in \emph{2014 IEEE 79th Vehicular Technology Conference (VTC
  Spring)}, 2014, pp. 1--5.

\bibitem{X.Du}
X.~{Du}, H.~{Van Nguyen}, C.~{Jiang}, Y.~{Li}, F.~R. {Yu}, and Z.~{Han},
  ``Virtual relay selection in lte-v: A deep reinforcement learning approach to
  heterogeneous data,'' \emph{IEEE Access}, vol.~8, pp. 102\,477--102\,492,
  2020.

\bibitem{Jabi}
M.~{Jabi}, M.~{Benjillali}, L.~{Szczecinski}, and F.~{Labeau}, ``Energy
  efficiency of adaptive harq,'' \emph{IEEE Transactions on Communications},
  vol.~64, no.~2, pp. 818--831, 2016.

\end{thebibliography}

\end{document}